\documentclass[prl,twocolumn,showpacs,preprintnumbers,amsmath,amssymb,superscriptaddress]{revtex4}
\usepackage{graphicx}
\usepackage{color}

\begin{document}

\title{Microscopic understanding of the orbital splitting and its tuning at oxide interfaces}
\author{Zhicheng Zhong}
\affiliation{Institute of Solid State Physics, Vienna University of Technology, A-1040 Vienna, Austria}
\author{Philipp Wissgott}
\affiliation{Institute of Solid State Physics, Vienna University of Technology, A-1040 Vienna, Austria}
\author{Karsten Held}
\affiliation{Institute of Solid State Physics, Vienna University of Technology, A-1040 Vienna, Austria}
\author{Giorgio Sangiovanni}
\affiliation{Institut f\"ur Theoretische Physik und Astrophysik, Universit\"at W\"urzburg, Am Hubland, D-97074 W\"urzburg, Germany}

\date{\today}

\begin{abstract}
By means of a Wannier projection within the framework of density functional theory, we are able to identify the modified $c$-axis hopping and the energy mismatch between the cation bands as the main source
 of the $t_{2g}$ splitting around the $\Gamma$ point for oxide heterostructures, excluding previously proposed mechanisms such as Jahn-Teller distortions or electric field asymmetries.
Interfacing LaAlO$_3$, LaVO$_3$, SrVO$_3$ and SrNbO$_3$ with SrTiO$_3$ we show how to tune this orbital splitting, designing heterostructures  with more  $d_{xy}$ electrons at the interface. 
Such an ``orbital engineering'' is the key for controlling the physical properties at the interface  of oxide heterostructures.
\end{abstract}

\maketitle

When an atom is part of a periodic arrangement, such as a solid, the spherical symmetry of its potential is lowered with respect to the case of a free atom.
Typical is the example of transition metal ions surrounded by oxygen octahedra in cubic perovskites.
Consequently, e.g., the five $d$ orbitals split: In bulk SrTiO$_3$ one has three $t_{2g}$ orbitals ($d_{xy}$, $d_{xz}$ and $d_{yz}$) and two $e_g$ ones ($d_{x^2-y^2}$ and $d_{3z^2-r^2}$). 
The physics of transition metal oxides is deeply influenced by the further (finer) splittings of the $t_{2g}$ or $e_g$ orbitals, for instance when distortions of the octahedra are energetically favored \cite{imadaRMP70}. 
Both from the point of view of basic materials research and from that of technological development and device fabrication, it would be fascinating if one could control such deviations from the perfect cubic perovskite structure by means of some external adjustable parameters.
Yet, this is something not easy to do in a flexible and controlled way in bulk materials.

The recent breakthrough in growing oxide heterostructures, such as LaAlO$_3$ grown on SrTiO$_3$, offers a new possibility to tune the orbital degrees
of freedom.
The interface breaks  the translational, and hence the cubic, symmetry.
As a consequence, the three $t_{2g}$ orbitals of the Ti atoms close to the interface split \cite{popovicPRL101}. 
Evidences for similar effects come also from experiments on a bare SrTiO$_3$ (001) surface upon cleavage \cite{santandersyroNature469,meevasanaNatureMat10}.
The physical properties of the entire heterostructure, such as superconductivity \cite{reyrenScience317}, phase separation \cite{ariandoNatComm2}, magnetism \cite{brinkmanNatMat,liNatPhys7,bertNatPhys7}, etc., depend on the $t_{2g}$ electrons at the interface \cite{singPRL102,salluzzoPRL102}.
These effects have been revealed in a series of stunning experiments \cite{ohtomoNature419,ohtomoNature427,reyrenScience317}, which also demonstrated that interfaces in layered oxide heterostructures can not only be seen as a way of tuning orbital splittings for specific desired purposes, but also for engendering new physical effects that are absent in the constituent bulk materials.

\begin{figure}[bh]
\includegraphics[width=7cm]{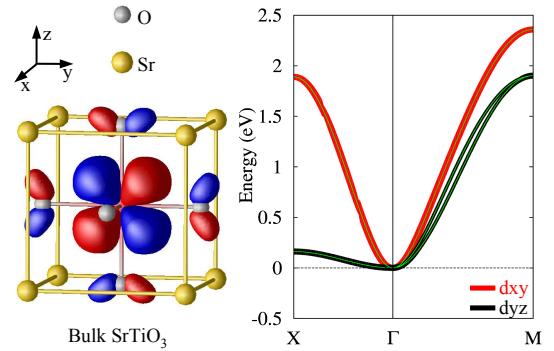}
\caption{Right panel: Band structure (green lines) for bulk SrTiO$_3$ of two of the three $t_{2g}$ bands: the $d_{xy}$ (red) and the $d_{yz}$ (black). The thickness is proportional to the orbital character. Left panel: Crystal structure of bulk SrTiO$_3$ and illustration of the maximally localized Wannier function with $d_{yz}$-character centered on the Ti site.} 
\label{Fig:first}
\end{figure}

A {\it sine qua non} condition for a successful ``orbital engineering'' in these layered systems is therefore the understanding of the mechanism behind the above-mentioned lifting of the $t_{2g}$ degeneracy.
Several density functional theory (DFT) calculations have been applied to LaAlO$_3$/SrTiO$_3$ (LAO/STO) heterostructures \cite{pentchevaPRB08,zhongPRB82,chenPRB82,leePRB78,pentchevaPRB74,mauricePSSA203,stengelPRL106,delugasPRL106}.
They clearly show that the $d_{xy}$ band with Ti character close to the interface is no longer degenerate with the $d_{xz}$ and $d_{yz}$ bands.
Some mechanisms for this effect have been proposed such as Jahn-Teller distortions \cite{mauricePSSA203}, a crystal field splitting because of the Sr$^{2+}$ vs.\ Ca$^{3+}$ asymmetry at the interface \cite{pentchevaPRB74,leePRB78} and a wedge-like or quantum-well-like potential at the interface \cite{santandersyroNature469, stengelPRL106}.
A simple, microscopic understanding of this mechanism is however hitherto lacking.

 In this Letter, we determine the relevant parameters governing the splitting between the $d_{xy}$ and $d_{xz}/d_{yz}$ orbitals at the interface by a Wannier function projection. We show that the $c$-axis hopping and cation energy mismatch 
is of primary importance,  ruling out other potential mechanisms.
This understanding also allows for a tailor-made orbital splitting, which in turn controls  the physical properties of oxide heterostructures.

In the right panel of Fig. \ref{Fig:first} we show the bands with $t_{2g}$ character for bulk SrTiO$_3$ as calculated by Wien2k in the paramagnetic phase employing the generalized gradient approximation (GGA) \cite{wien2k}.
The bulk structure is perfectly cubic, therefore the three bands are degenerate at the $\Gamma$-point. 
Along the $\Gamma$-X direction the $xy$ and $xz$ (not shown here) bands are strongly dispersive, due to the large hopping amplitude along the $x$-direction, while the $yz$ orbital (shown in the left part of the Figure) overlaps much less in that direction. The values of the hoppings are listed in the Table in the Supplementary Material. 

\begin{figure}[tb]
\includegraphics[width=8cm]{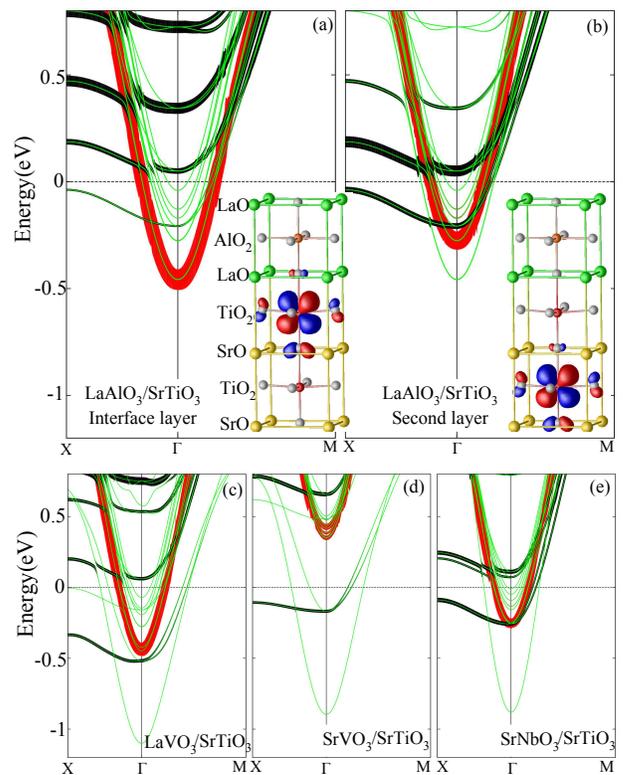}
\caption{(a) and (b): Band structure for the LaAlO$_3$/SrTiO$_3$ heterostructure with one AlO$_2$ layer symmetrically surrounded by two LaO layers and STO. The thickness of the lines highlights the character of (a) the interface layer, (b) the second TiO$_2$ layer. Shown are also the Wannier orbitals for the two layers.
(c)-(e): Band structure for LaVO$_3$/ (c), SrVO$_3$/ (d) and SrNbO$_3$/SrTiO$_3$ (e). The lowest parabolic bands (thin green line) are almost completely of V or Nb $d_{xy}$ character, panel (c)/(d) and (e), respectively.
In all panels red (black) indicates ${xy}$ (${yz}$) character. 
}
\label{Fig:second}
\end{figure}

A very simple way of addressing the question of the $t_{2g}$ splitting is to build a symmetric heterostructure with (for the case of LAO/STO) one AlO$_2$ layer in the middle, two LaO layers around it and alternating TiO$_2$ and SrO layers on both sides (see Fig. \ref{Fig:second}).
Heterostructures of this kind have two $n$-type of interfaces: TiO$_2$$\mid$LaO and LaO$\mid$TiO$_2$. This implies that the interface will always be metallic, mimiking the situation of an $n$/$p$ heterostructure above the critical thickness.
The calculations have been done using the Wien2k code \cite{wien2k} with GGA and optimizing the internal coordinates. 

Several groups have analyzed the band structure of LAO/STO. The picture coming out of the DFT calculation and supported by a number of experimental findings, suggests the presence of two types of carriers: a more itinerant one extending deep into the STO part and one more confined at the interface, primarily made of $d_{xy}$ electrons \cite{popovicPRL101,santandersyroNature469}.
The former is easier to get trapped by impurities, defects or in the form of lattice polarons. 
Therefore, by tuning the splitting at the $\Gamma$ point between the $d_{xy}$ and $d_{yz}$ bands we can control these two components and, in turn, the metallic or insulating behavior of different heterostructures.
We consider here different heterostructures and for each of them we calculate the band structure focusing on the orbital character close to the interface.

A very transparent description can be achieved by extracting Wannier functions of $d$ character and calculating the tight-binding Hamiltonian arising from them.
To our knowledge, this has been never done in the context of the heterostructures but it is actually a very powerful and simple way of describing the electronic properties of the different cations at, as well as close to, the interface.
In our case we consider maximally localized Wannier functions, because the resulting low-energy Hamiltonian can then also be very naturally used to perform future many-body calculations. 
The Wannier projection was performed with the Wien2Wannier package \cite{kunesCPC181} interfacing Wien2k to Wannier90 \cite{wannier90}.

The two pieces of information that we are going to focus on are $i$) the local term of the Hamiltonian, which gives the energy position of a given orbital (or, more precisely of the center of gravity of the corresponding band) and $ii$) the nearest-neighbor hopping terms between different $t_{2g}$ orbitals, parallel and perpendicular to the interface plane.
From the projection onto maximally localized Wannier functions we directly get both.
An analysis of the onsite energies for the different heterostructures allows us to identify the mechanism causing the shift of the different bands: the polar discontinuity between LaO$^{(+)}$ and SrO$^{(0)}$ layers or the presence of electrostatic potential shift induced by the accumulation of charge at the interface, or possible octahedra distortions \cite{stengelPRL106,delugasPRL106}.
The hopping amplitudes between the Wannier orbitals give us information about the splitting at the $\Gamma$-point and about the degree of itinerancy of each orbital, another point often discussed in these ``subband'' studies \cite{popovicPRL101,santandersyroNature469}. 
From our Wannier functions we can also analyze the localization of the orbitals in real-space as well as establish the amount of orbital reconstruction induced by the presence of the interface.

In Fig.~\ref{Fig:second}a and \ref{Fig:second}b the band structure for the $n$/$n$ LAO/STO heterostructure is shown.
Even in the presence of the interface, the Wannier functions are very close to ideal $t_{2g}$ orbitals. Therefore in Fig. \ref{Fig:second} we only show the $d_{yz}$ one (the $d_{xz}$ is symmetrically related), for which a small degree of asymmetry can be observed.
Unlike for the bulk, we observe a splitting of the $t_{2g}$ bands at the interface, in agreement with earlier calculations \cite{popovicPRL101,santandersyroNature469}: The bottom of the $d_{xy}$ band orbital at the interface is lower than that of the lowest $d_{yz}$.
In addition to that, the bands with a component in the $z$ direction (of which only the $d_{yz}$ is shown here) give rise to several branches close to the interface, which are due to the quantum confinement effect along the $z$ direction. 

\begin{figure}[th]
\includegraphics[width=8cm]{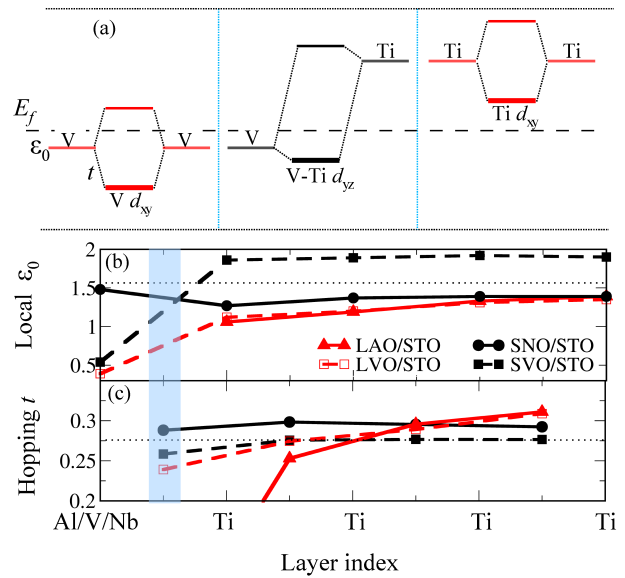}
\caption{(Color online) Panel a: level scheme for SVO/STO for in-plane (V-V or Ti-Ti) and out-of-plane (V-Ti) hopping processes. Panel b and c: Layer dependence of the average over the three $t_{2g}$ bands of, respectively, the on-site energy and of the $d_{yz}$-$d_{yz}$ hopping amplitude the along the $z$ direction. The dashed lines represent the bulk values.}
\label{Fig:third}
\end{figure}

The major advantage of our Wannier projection is that we can now identify the mechanism behind the 250meV $d_{xy}$ \emph{vs.} $d_{yz}$ splitting observed at the $\Gamma$ point. Indeed, from our single-particle Hamiltonian in the Wannier basis, we can reliably extract the local terms $\varepsilon_0^{xy}$ and $\varepsilon_0^{yz}$. 
First, we can quantify the band bending at the interface: As one can see in Fig.~\ref{Fig:third}b this amounts roughly to 0.3 eV across three layers, in agreement with experimental indications \cite{yoshimatsuPRL101} as well as with previous DFT calculations \cite{delugasPRL106,janickaPRL102}.
Second, and even more important, since the difference between $\varepsilon_0^{xy}$ and $\varepsilon_0^{yz}$ is as small as 50meV at the interface (see Supplementary Material) we can unambiguously rule out the polar discontinuity as the main source for the $d_{xy}$ \emph{vs.} $d_{yz}$ splitting observed at the $\Gamma$ point.

A second effect determining the $d_{xy}$ \emph{vs.} $d_{yz}$ splitting at the $\Gamma$ point is the following:
When LAO is grown on STO a strong reduction of the vertical $d_{yz}$-$d_{yz}$ and $d_{xz}$-$d_{xz}$ hoppings occurs due to the presence of the insulating LAO overlayer (a similar mechanism has been already discussed for Ni-based heterostructures \cite{chaloupkaPRL100}). 
The hopping amplitude along the $z$-axis involving the $d_{xy}$ orbital is instead small both for bulk STO and for the LAO/STO (see Table in the Supplementary Material). 
Since only the $d_{yz}$ and $d_{xz}$ hopping processes along $z$ are affected the degeneracy with the $d_{xy}$ is lifted and the bottom of the $d_{yz}$ and $d_{xz}$ bands is shifted up (green/black band in Fig.~\ref{Fig:second}a).
Quantitatively, this second effect coming from the hopping reduction dominates:
While the difference between the ${xy}$ and the ${yz}$ local terms is about 50meV, the suppression of the hopping along the $z$ direction accounts for the remaining $\sim$200meV inband splitting, bringing the total splitting at the $\Gamma$ point to the correct value which can be read off from Fig. \ref{Fig:second}a.

Now that we have the two ``control knobs'', the reduction of $c$-axis hopping and the energy position of the $t_{2g}$ bands of the different cations, we can use them to engineer the $d_{xy}$ \emph{vs.} $d_{yz}$ splitting at the $\Gamma$ point back to the vanishing bulk value or even to reverse it, pushing the $d_{xy}$ orbital above the $d_{yz}$. 
Since the biggest effect is coming from the $c$-axis hopping reduction we first focus on the case of a heterostructure in which the polar discontinuity given by the overlayer is similar to that of LAO/STO but in which the hopping along the $z$ direction is large.  
This way we expect to bring the bottom of the $d_{yz}$ band down in energy.
Such a situation is achieved in the case of LaVO$_3$/SrTiO$_3$ (LVO/STO), shown in Fig.~\ref{Fig:second}c. 
The bottom parabolic band (thin green line) is of V ${xy}$ character. However, now the band with $d_{yz}$ character is a Ti-V hybrid.
Like in LAO/STO, the LaO and VO$_2$ layers in LVO/STO have, respectively, nominal charge +1 and -1, but the presence of $d$ electrons on V now lets the $yz$ hopping survive. 
In Fig.~\ref{Fig:second}c we indeed see that, at the interface, the $d_{xy}$ band has been pushed above the lowest (Ti-V hybridized) $d_{yz}$ branch at the $\Gamma$ point, giving a situation totally different from LAO/STO.

As next ``orbital tweak'' we want to push the $d_{xy}$ orbital further up in energy in order to design an interface with $d_{yz}$/$d_{xz}$ electrons only.
This can be done by replacing La with Sr, i.e. by growing SrVO$_3$/SrTiO$_3$ (SVO/STO) \cite{yoshimatsuPRL104}.
Indeed, since the difference between the V and Ti Wannier local levels in the heterostucture is about $1.3$eV (see Fig. \ref{Fig:third}b and Table in the Supplementary Material) we expect the situation illustrated in the scheme shown in Fig.~\ref{Fig:third}a:
The Ti $d_{xy}$ band is higher in energy and a mixed Ti/V band with $d_{yz}$ character is formed. 
Both effects can be observed in the actual calculation shown in Fig.~\ref{Fig:second}d, where the Ti/V $d_{yz}$ band crossing the Fermi level is represented by the green and thicker black lines.
Such Ti/V $d_{yz}$ band comes from the fact that, unlike the $d_{xy}$ orbital, the $d_{yz}$ and $d_{xz}$ orbitals of Ti and V hybridize to each other along the $z$-axis at the interface.
The same hybridization is also responsible for the $d_{xy}$ \emph{vs.} $d_{yz}$ splitting of the V states: The $d_{yz}$ orbital is much higher in Ti than in V and this ``pushes'' the V $d_{yz}$ orbital up in energy, as shown in the level scheme of Fig.~\ref{Fig:third}a.  

\begin{figure}[bt]
\includegraphics[width=8cm]{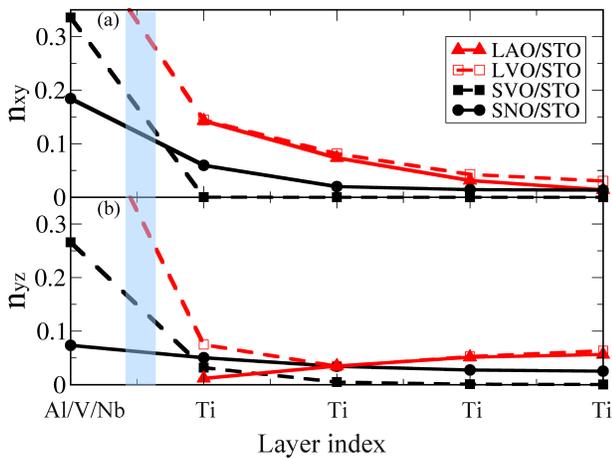}
\caption{(Color online) Layer dependence of carrier densities of $d_{xy}$ (a) and $d_{yz}$ (b) orbitals}
\label{Fig:fourth}
\end{figure}

We have therefore designed a heterostructure with $d_{yz}$ (and $d_{xz}$) carriers only at the interface. 
To achieve this, the big energy mismatch between Ti and V $t_{2g}$ orbitals in SVO/STO has been exploited. 
A case in which such mismatch is instead as small as 0.2eV is the SrNbO$_3$/SrTiO$_3$ heterostructure (SNO/STO).
As a result, the splitting around $\Gamma$ looks close to that of an undistorted bulk perovskite with three degenerate $t_{2g}$ bands.
In other words with SNO/STO we can tune the $t_{2g}$ splitting in such a wat that we get a seemingly ``accidental'' degeneracy as in the bulk case. 
Considering also the contribution from the hopping terms we can account for the $\sim$0.7eV difference in the position of the Nb $d_{xy}$ and Nb-Ti hybridized $d_{yz}$ bands observed in Fig.~\ref{Fig:second}e.

The ability of tuning the orbital splitting at the $\Gamma$ point of oxide interfaces is not only {\it per se} intriguing but it also guides us to make heterostructures with desired properties: insulating, metallic, with more or less localized carriers at the interface, etc. 
To illustrate this, we show the $xy$ and $yz$ occupancies in Fig. \ref{Fig:fourth} for the four heterostructures considered here (the behavior of the hopping terms and of the local levels is shown instead in Fig.~\ref{Fig:third}b and c respectively and the values are reported in the Supplementary Materials).
In the lower panel of Fig. \ref{Fig:fourth} the $yz$ component is shown. This extends into the STO part and, since it has a large degree of itinerancy it is the component which gets more easily trapped hence it hardly contributes to transport (i.e. it is not observed in Hall measurements).
The upper panel shows instead the $xy$ electrons, a part of which forms a markedly two-dimentional electron liquid at the interface.
Evidently, SVO/STO has no Ti $xy$ carriers with high mobility.
SNO/STO has a slightly larger fraction of $d_{xy}$ electrons and LAO/STO has more. 
On the basis of this argument, the most metallic heterostructure should be LVO/STO. Yet, since V has two $d$-electrons in LaVO$_3$, strong correlation effects can contribute to localize them, thereby reducing the number of carriers available for transport.
Experimentally, LAO/STO and LVO/STO with polar discontinuity have shown conductivity with highly mobile carriers \cite{ohtomoNature427,hottaPRL99,mueller_unpublished}. 
Our theoretical study suggests that SNO/STO will have similar transport properties without polar discontuinity, whereas SVO/STO will have no $xy$ carriers on the Ti interface layer.

In conclusion, we have extracted maximally localized Wannier functions for four different heterostructures. This allows for a micoscopic understanding on how 
 to tune the  orbital degrees of freedom at the interface. 
The $d_{xy}$ \emph{vs.} $d_{yz}$ splitting at the $\Gamma$ point is mainly determined by the hopping amplitude along the $z$ direction which is strongly influenced by the presence of the interface as well as by the energy difference between the Ti bands and those of the capping cations. 
We also showed that upon changing the material grown above STO we can tune this hopping and the shift of the local levels in such a way that the amount of $d_{xy}$ electrons can be engineered in a tailor-made way. 
This can be exploited to increase the amount of $d_{xy}$ electrons and, hence, to make more metallic heterostructures.

We thank E.~Assmann, P.~Blaha, R.~Claessen and M.~Sing for valuable discussions. Z.Z. acknowledges financial support from the EU-Indian network MONANI, P.W. from the Austrian Science Fund (FWF) through SFB ViCom F4103-N13 and K.H. through the research unit DFG FOR-1346/FWF I597-N16.


\begin{thebibliography}{99}
\bibitem{imadaRMP70} M.~Imada, A.~Fujimori and Y.~Tokura, Rev. Mod. Phys. {\bf 70}, 1039 (1998)
\bibitem{popovicPRL101} Z.~S.~Popovic, S.~Satpathy and R.~M.~Martin, Phys. Rev. Lett. {\bf 101}, 256801 (2008)
\bibitem{santandersyroNature469} A.~F.~Santander-Syro, {\it et al.}, Nature {\bf 469}, 189 (2011)
\bibitem{meevasanaNatureMat10} W.~Meevasana, P.~D.~C.~King, R.~H.~He, S-K.~Mo, M.~Hashimoto, A.~Tamai, P.~Songsiriritthigul, F.~Baumberger and Z-X.~Shen, Nature Mat. {\bf 10}, 114 (2011)
\bibitem{reyrenScience317} N.~Reyren, S.~Thiel, A.~Caviglia, L.~Fitting Kourkoutis, G.~Hammerl, C.~Richter, C.~W.~Schneider, T.~Kopp, A.-S.~R\"uetschi, D.~Jaccard, M.~Gabay, D.~A.~Muller, J.-M. Triscone and J.~Mannhart, Science {\bf 317}, 1196 (2007).
\bibitem{ariandoNatComm2} Ariando, X.~Wang, G.~Baskaran, Z.~Q.~Liu, J.~Huijben, J.~B.~Yi, A.~Annadi, A.~Roy~Barman, A.~Rusydi, S.~Dhar, Y.~P.~Feng, J.~Ding, H.~Hilgenkamp and T.~Venkatesan, Nature Comm. {\bf 2}, 188 (2011)
\bibitem{brinkmanNatMat} A.~Brinkman, M.~Huijben, M.~van~Zalk, J.~Huijben, U.~Zeitler, J.~C.~Maan, W.~G.~van~der~Wiel, G.~Rijnders, D.~H.~A.~Blank and H.~Hilgenkamp {\bf 6}, 493 (2007)
\bibitem{liNatPhys7} Lu Li, C.~Richter, J.~Mannhar and R.~C.~Ashoori, Nature Physics {\bf 7}, 762 (2011)
\bibitem{bertNatPhys7} J.~A.~Bert, B.~Kalisky, C.~Bell, M.~Kim, Y.~Hikita, H.~Y.~Hwang and K.~A.~Moler, Nature Physics {\bf 7}, 767 (2011)
\bibitem{singPRL102} M.~Sing, G.~Berner,  K.~Go\ss{}, A.~ M\"uller, A.~Ruff, A.~Wetscherek, S.~Thiel, J.~Mannhart, S.~A.~Pauli, C.~W.~Schneider, P.~R.~Willmott, M.~Gorgoi, F.~Sch\"afers and R.~Claessen, Phys. Rev. Lett. {\bf 102}, 176805 (2009)
\bibitem{salluzzoPRL102} M.~Salluzzo, J.~C.~Cezar, N.~B.~Brookes, V.~Bisogni, G.~M.~De~Luca, C.~Richter, S.~Thiel, J.~Mannhart, M.~Huijben, A.~Brinkman, G.~Rijnders and G.~Ghiringhelli,  Phys. Rev. Lett. {\bf 102}, 166804 (2009)
\bibitem{ohtomoNature419} A.~Ohtomo, D.~A.~Muller, J.~L.~Grazul and H.~Y.~Hwang, Nature {\bf 419}, 378 (2002)
\bibitem{ohtomoNature427} A.~Ohtomo and H.~Y.~Hwang, Nature {\bf 427}, 423 (2004)
\bibitem{pentchevaPRB08} R.~Pentcheva and W.~Pickett, Phys. Rev. B {\bf 78}, 205106 (2008)
\bibitem{zhongPRB82} Zhicheng~Zhong, P.~X.~Xu and P.~J.~Kelly, Phys. Rev. B {\bf 82}, 165127 (2010)
\bibitem{chenPRB82} H.~Chen, A.~Kolpak and S.~Ismail-Beigi, Phys. Rev. B {\bf 82}, 085430 (2010)
\bibitem{leePRB78} J.~Lee and A.~Demkov, Phys. Rev. B {\bf 78}, 193104 (2008) 
\bibitem{pentchevaPRB74} R.~Pentcheva and W.~Pickett, Phys. Rev. B {\bf 74}, 035112 (2006)
\bibitem{mauricePSSA203} J.-L.~Maurice, C.~Carretero, M.-J.~Casanove, K.~Bouzehouane, S.~Guyard, \'E.~Larquet, J.-P.~Contour, Physica Status Solidi A {\bf 203}, 2209 (2006)
\bibitem{stengelPRL106} M.~Stengel, Phys. Rev. Lett. {\bf 106}, 136803 (2011) 
\bibitem{delugasPRL106} P.~Delugas, A.~Filippetti, V.~Fiorentini, D.~I.Bilc, D.~Fontaine and P.~Ghosez, Phys. Rev. Lett. {\bf 106}, 166807 (2011)
\bibitem{wien2k} P. Blaha {\it et al.}, {\it WIEN2k, An Augmented Plane Wave Local Orbitals Program for Calculating Crystal Properties}, edited by Karlheinz Schwarz (Technische Universit\"at Wien, Austria, 2001). 
\bibitem{kunesCPC181} J. Kune\v{s}, R. Arita, P. Wissgott, A. Toschi, H. Ikeda, and K. Held, Comp. Phys. Comm. {\bf 181}, 1888 (2010)
\bibitem{wannier90} A.~Mostofi, J.~Yates, Y.-S.~Lee, I.~Souza, D.~Vanderbilt and N.~Marzari, Comp. Phys. Comm. {\bf 178}, 685 (2008).
\bibitem{yoshimatsuPRL101} K.~Yoshimatsu, R.~Yasuhara, H.~Kumigashira and M.~Oshima Phys. Rev. Lett. {\bf 101}, 026802 (2008)
\bibitem{janickaPRL102} K.~Janicka, J.~Velev and E.~Tsymbal, Phys. Rev. Lett. {\bf 102}, 106803 (2009)
\bibitem{chaloupkaPRL100} J.~Chaloupka and G. Khaliullin, Phys. Rev. Lett. {\bf 100}, 016404 (2008)
\bibitem{yoshimatsuPRL104} K.~Yoshimatsu, T.~Okabe, H.~Kumigashira, S.~Okamoto, S.~Aizaki, A.~Fujimori and M.~Oshima, Phys. Rev. Lett. {\bf 104}, 147601 (2010)
\bibitem{hottaPRL99} Y.~Hotta, T.~Susaki and H.~Y.~Hwang, Phys. Rev. Lett. {\bf 99}, 236805 (2007)
\bibitem{mueller_unpublished} A.~M\"uller, {\it et al.}, unpublished
\end{thebibliography}
\end{document}